\documentstyle[12pt,epsfig]{article}
     \def\lsim{\raise0.3ex\hbox{$<$\kern-0.75em\raise-1.1ex\hbox{$\sim$}}}
\def\gsim{\raise0.3ex\hbox{$>$\kern-0.75em\raise-1.1ex\hbox{$\sim$}}}
\def\noi{\noindent}
\def\nn{\nonumber}
\def\bea{\begin{eqnarray}}  \def\eea{\end{eqnarray}}
\def\beq{\begin{equation}}   \def\eeq{\end{equation}}

\def\beeq{\begin{eqnarray}} \def\eeeq{\end{eqnarray}}

\begin{document}
\begin{center}
{\Large \bf }\par 
\vskip 3 truemm
  {\Large \bf ${\bf J/}\psi$ suppression and the nuclear absorption } \vskip 3 truemm
  {\Large \bf decrease with increasing energy}
   \vskip 1 truecm
{\bf A. Capella$^1$ and E. G. Ferreiro$^{2,3}$}\\ 
$^1$Laboratoire de Physique Th\'eorique
UMR 8627 CNRS,
Universit\'e de Paris XI, F-91405 Orsay cedex, France\\
$^2$Departamento de F{\'\i}sica de Part{\'\i}culas, 
Universidad de Santiago de Compostela, 15782 Santiago de Compostela, 
Spain\\
$^3$Laboratoire Leprince-Ringuet, Ecole Polytechnique-IN2P3/CNRS, F-91128 Palaiseau cedex, France

\end{center}
\vskip 0.5 truecm
\begin{abstract}
In the field theoretical Glauber-Gribov framework in the eikonal approximation, we show that, in hadron-nucleus collisions in the high-energy limit and under specific conditions, the rescattering contributions to the $J/\psi$ production cross-section vanish, leading to an $A^1$ dependence on the atomic number $A$ -- except for nuclear effects in the nucleus structure function, the so-called shadowing. 
We show that the RHIC data on $J/\psi$ production in $dAu$ collisions can be described with nuclear shadowing alone $(\sigma_{abs} = 0$). The corresponding results for $Au$ $Au$ collisions are also presented and compared with available data.
\end{abstract}

\vskip 1 truecm
\noi LPT-Orsay-06-42\par
\newpage
\pagestyle{plain} 

\section{Introduction}

A surprising result in $J/\psi$ production in hadron-nucleus (or deuteron-nucleus) collisions is the substantial decrease of the so-called nuclear absorptive cross-section, $\sigma_{abs}$, between CERN 
and RHIC energies \cite{1r,2r}.
In this paper we propose a scenario which explains the observed decrease of $\sigma_{abs}$ with increasing $s$. \par

We are going to proceed as follows: 
In Section 2 we develop our formalism for the absorptive cross-section $\sigma_{abs}$
in the framework of Glauber-Gribov field theory. 
In Section 3 we introduce a dynamical, non linear nuclear shadowing. It is a consequence of multiple scattering, 
and it is determined in terms of diffractive cross sections.
In Section 4 
we present our results for 
$dAu$ and $AuAu$ collisions at RHIC energies.
We will finish by our conclusions in Section 5.

\section{The absorptive cross-section}

We proceed in the framework of Glauber-Gribov \cite{3r} field theory in the eikonal approximation. We shall consider rescattering of both the 
light partons of the projectile, with a typical hadronic cross-section $\sigma$, and the heavy $c\overline{c}$ system, 
with cross-section $\widetilde{\sigma}$. We shall assume that the dominant contribution for the $J/\psi$
production 
 is the conventional mechanism 
such as gluon-gluon fusion. In the field theoretical language, 
we assume that the $J/\psi$ is produced in a cut interaction of the projectile with a target nucleon. 
Production in the blob of a diagram is assumed to be small.\par

For simplicity we shall assume throughout this paper that the amplitudes of both the projectile and heavy system scattering are purely imaginary. 
%
%
%
Let us denote by $\sigma_A^{in} (\sigma)$ the $pA$ inelastic cross-section, at fixed impact parameter,
calculated in the Glauber model with $\sigma_{pN} = \sigma$, 
i.e.\footnote{For simplicity we have used the notation cross-section for this 
quantity. In fact it corresponds to a dimensionless quantity, the 
differentiated
cross-section $d\sigma/d^2b$. This is also valid for the later equations, where 
we have used this kind of notation too.}
\beq
\label{1e}
\sigma_A^{in} (\sigma ) =  1 - \left [ 1 - \sigma \ T_A(b)\right ]^A =
\sum_{k=1}^A {A \choose k} \sigma^k  
T_A^k(b) \left [ 1 - \sigma T_A(b) \right ]^{A-k}\ .
\eeq

\noi Here $T_A(b)$ is the nuclear profile function normalized to unity,
and $k$ is the number of cut interactions. 
Since we allow interactions of both the projectile, with cross-section 
$\sigma$, and the heavy system, with cross-section $\widetilde{\sigma}$, to take place, the cross-section
of the $J/\psi$ production is proportional to $\sigma_A^{tot}(\sigma + \widetilde{\sigma})$ \cite{7r}.
However, since we require that at least one interaction is cut, only its inelastic part 
$\sigma_A^{in} (\sigma + \widetilde{\sigma})$ does contribute. Moreover, in our mechanism the $J/\psi$
is produced in a cut interaction of the projectile. 
Thus, we have to substract from $\sigma_A^{in} (\sigma + \widetilde{\sigma})$ the contribution with no cut 
interaction of the projectile, i.e.
\beq
\label{2e}
\sum_{k=1}^A {A \choose k} \left [ (\sigma+ \widetilde{\sigma})^k - \widetilde{\sigma}^k 
\right ] T_A^k(b) \left ( 1 - (\sigma + \widetilde{\sigma})T_A(b) \right )^{A-k}
= \sigma_A^{in} (\sigma )\ .
\eeq
Without the second term in the squared bracket (proportional to $\widetilde{\sigma}^k$) we have just the 
expression of $\sigma_A^{in} (\sigma+ \widetilde{\sigma})$. This second term has been substracted since at least
one cut interaction of the projectile (in which the heavy system is produced) has to be present. We see
from (\ref{2e}) that 
this difference is equal to $\sigma_A^{in} (b)$. This means that
all eikonal type rescatterings of the heavy system have cancelled out.\par

The origin of this result is a cancellation between different discontinuities (cuttings) of the diagrams in the Glauber-Gribov theory which is contained in the eikonal model -- but has a more general validity. \par

Let us now decompose $\sigma$ into two terms, a small one corresponding to the production of $J/\psi$ and the rest in which the $J/\psi$ is not produced~: $\sigma = \sigma_{pN}^{\psi} + \sigma_{pN}^{N_0\psi}$. 
From eq. (\ref{2e}), it is clear that the contribution to the $J/\psi$ cross-section can be written as 
\beq
\label{3e}
\sigma_{pA}^{\psi} (b) = \sum_{k=1}^A {A \choose k} \left [ \sigma^k - (\sigma_{pN}^{N_0\psi})^k \right ]\ T_A^k(b)\left [ 1 - T_A(b) \sigma ) \right ]^{A-k} = \sigma_A^{in} (\sigma_{pN}^{\psi}) \sim A \sigma_{pN}^{\psi} \ T_A (b)\ .
\eeq

\noi We see that the same cancellation is at work. 
We see from eq. (\ref{3e}) that
only rescatterings of eikonal type involving the small piece, $\sigma_{pN}^{\psi}$, of $\sigma$ are left 
(self absorption) \cite{4r}.
%
The conventional shadowing, related to diffractive cross-section $\sigma_D$, 
is due to diagrams of a non-eikonal type involving the triple Pomeron coupling and it is not included in the above 
formulae. It can be easily incorporated, as we will do below.\par

Apart from these conventional shadowing corrections, we see from eq. (\ref{3e}) that $J/\psi$ production is proportional to $A^1$. We shall show below that this result is valid only at high energy. It is in sharp contrast with the situation at low energies where a rescattering of the pre-resonant $c\overline{c}$ system with the target nucleons takes place, with cross section $\sigma_{abs} = (1 - \varepsilon ) \widetilde{\sigma}$.
$\sigma_{abs}$ is the cross-section of the interaction of the $c\overline{c}$ pair with a target nucleon in which the $c\overline{c}$ is transformed in such a way that it has no projection into $J/\psi$ (open charm is produced instead). We can write $\sigma_{abs} = (1 - \varepsilon ) \widetilde{\sigma}$ where $\widetilde{\sigma}$ is the total $c\overline{c}-N$ cross-section and $\varepsilon \widetilde{\sigma}$ ($\varepsilon < 1$) is the contribution to $\widetilde{\sigma}$ of all intermediate states having a non-zero projection into $J/\psi$. Since this contribution is expected to be quite small we should have $\varepsilon \ll 1$ for $J/\psi$ and, on the contrary, $\varepsilon \sim 1$ for open charm production. Thus, for open charm $\sigma_{abs} = (1 - \varepsilon ) \widetilde{\sigma} \sim 0$ while for $J/\psi$ production $\sigma_{abs} \sim \widetilde{\sigma}$.  
\par

The origin of the change in the physical picture between low and high energies is well known \cite{5r} \cite{6r} \cite{7r}. The mass $M_{c\overline{c}}$ of the heavy system introduces a new scale, the coherence length $\ell_c$
\beq
\label{4e}
\Delta \equiv {1 \over \ell_c} = m_p {M_{c\overline{c}}^2 \over s\ x_1} \ .
\eeq

\noi At high energies the coherence length is large and the projectile interacts with the nucleus as a whole. At low energies the coherence length is small and the probabilistic picture with a longitudinal ordering in $z$ of the various interactions is valid. In the first interaction at $z$ the heavy system is produced and in successive interactions, at larger values of $z$, it rescatters with nucleons along its path. The change with energy of the rescattering mechanism has a clear physical interpretation. Due to the presence of the heavy system some contributions to the production cross-section (corresponding to some particular discontinuity of a given diagram) contain a non-vanishing minimal momentum transfer ($t_{min} \not= 0$) and are suppressed by the nuclear form factor.\par

Following ref. \cite{7r} the effects of the coherence length can be taken into account by writing the total $pA$ cross-section as
\beq
\label{5e}
{\sigma}_A^{tot} (b) = \sum_{n=1}^A {A \choose n} \sum_{j=1}^n T_n^{(j)}(b) \sigma_n^{(j)} 
\eeq 

\noi where $n$ is the number of interactions (both cut and uncut) of both the projectile and the heavy system. Here we have to consider a particular ordering of the longitudinal coordinates $z_i$ of the $n$ interacting nucleons~: $z_1 \leq z_2 \leq \cdots \leq z_n$,  and the $n$-th power of the nuclear profile function $T_A^n (b)$ is replaced by 
\beq
\label{6e}
T_n^{(j)}(b) = n! \int_{-\infty}^{+ \infty} dz_1 \int_{z_1}^{+ \infty} dz_2 \cdots \int_{z_{n-1}}^{+ \infty} dz_n\
\cos ( \Delta (z_1 - z_j) )\ \prod_{i=1}^n \rho_A (b, z_i)
\eeq

\noi where $\Delta$ is the inverse of the coherence length, eq. (\ref{4e}), and $\rho_A$ is the nuclear density. Note that for $\Delta = 0$, corresponding to asymptotic energies, all integrals $T_n^{(j)}(b)$ are equal to $T_A^n(b)$. For $\Delta \not= 0$, the value of $j$ in eq. (\ref{5e}) depends on the discontinuity of the diagrams we are considering. Namely, interactions with nucleons 1 to $j-1$ are uncut and located to the left of the cutting line. Interaction $j$ can be 
either cut or located to the right of the cutting line. All other interactions from $j+1$ to $n$ can be cut or located to the right or to the left of the cutting line -- all possibilities have to be added. To this contribution we have to add its complex conjugate, which gives just a factor of 2
for purely imaginary amplitudes.\par

It is now straightforward to obtain $\sigma_n^{(j)}$ in eq. (\ref{5e}). We have to remember that when an interaction is uncut (located to either the right or the left of the cutting line), its amplitude ($ia = - \sigma /2$, for purely imaginary amplitudes) is unchanged. When the cutting takes place through the interaction, its amplitude 
is replaced by $\varepsilon \widetilde{\sigma}$ in the case of an interaction of the heavy system. Indeed, only the fraction $\varepsilon \widetilde{\sigma}$ of the $c\overline{c}-N$ cross-section, corresponding to the intermediate states with non-zero projection into $J/\psi$, does 
contribute\footnote{The introduction of $\varepsilon$ is most important at low energies where
the heavy system undergoes successive interactions with nucleons in its path and has to survive all of them. It does not
appear in the formula for $s \rightarrow \infty$.}. 
In the case of an interaction of the projectile its amplitude is replaced by $\sigma$. However, in order to keep track of the cut interactions of the projectile, we shall replace it by $\delta \sigma$ with $\delta = 1$. \par

Proceeding in this way it is easy to obtain the expressions for $\sigma_n^{(j)}$ in eq. (\ref{6e}) \cite{7r}. We get for $j > 1$ 
\beq
\label{7e}
\sigma_n^{(j)} = 2 \left ( - {\sigma \over 2} - {\widetilde{\sigma} \over 2} \right )^{j-1} \left [ {\sigma \over 2} +  {\widetilde{\sigma} \over 2} + \sigma (\delta-1) + \widetilde{\sigma} (\varepsilon - 1) \right ] \left ( \sigma (\delta - 1) + \widetilde{\sigma} (\varepsilon - 1)\right )^{n-j}
\eeq

\noi and for $j = 1$
\beq
\label{new1e}
\sigma_n^{(1)} = \left ( \sigma \delta + \widetilde{\sigma} 
\varepsilon \right ) \left ( \sigma (\delta - 1) + \widetilde{\sigma} (\varepsilon - 1)\right )^{n-1}\ .
\eeq

\noi The first factor of eq. (\ref{7e}) corresponds to the $j-1$ uncut interactions. The second one corresponds to the $j$th interaction which is either cut (which gives $\delta \sigma$ for the interaction of the projectile and $\varepsilon \widetilde{\sigma}$ for that of the heavy system) or located to the right of the cutting line (which gives $-(\sigma + \widetilde{\sigma})/2$). The last factor corresponds to the interactions $j+1$ to $n$. For $j=1$ the 
formula is different. Since a diagram can only be cut either through or in between interactions, the first interaction has to be cut and there is no complex conjugate term -- because all other interactions have to be located to the right of the cutting line.\par

In the high energy limit $s \to \infty$, $\Delta \to 0$ and $T_n^{(j)}(b) = T_A^n (b)$. It is easy to see that for $j > 1$ the last two terms in the square bracket of $\sigma_n^{(j)}$ cancel with the first two terms of $\sigma_n^{(j-1)}$. We are thus left with the first two terms of $\sigma_n^{(n)}$ the last two terms of $\sigma_n^{(2)}$ and $\sigma_n^{(1)}$. This gives
\beq
\label{8e}
{\sigma}_A^{tot} (b) = \sigma_A^{tot} (\sigma +  \widetilde{\sigma}) - \sigma_A^{in} \left ( \sigma (1 - \delta ) +  \widetilde{\sigma} (1 - \varepsilon )\right ) \ .
\eeq

\noi Since $\delta = 1$ the last term does not contain any collision of the projectile.
Only a part of the first term corresponds to the $J/\psi$ production mechanism we
have assumed.
More precisely, we have to consider only the inelastic part of $\sigma_A^{tot}(\sigma +  \widetilde{\sigma})$.
Moreover, as in eqs. (\ref{2e}) and (\ref{3e}), we have to remove the terms in which only $\widetilde{\sigma}$
and/or $\sigma^{N_0\psi}$ are present. We thus have
\bea
\label{9e}
\sigma_{pA}^{\psi} (b) &=& \sum_{k=1}^A
{A \choose k} \left [ (\sigma +  \widetilde{\sigma})^k - \left ( \sigma_{pN}^{N_0\psi} +\widetilde{\sigma} \right
)^k  \right ] T_A^k (b)\nn \\
&& \left [ 1 - (\sigma +  \widetilde{\sigma}) T_A(b) \right ]^{A-k} = \sigma_{A}^{in} (\sigma_{pN}^{\psi}) \simeq
A \sigma_{pN}^{\psi} T_A(b)
\eea

\noi which reproduces the result in eq. (\ref{3e}). \par

Let us now turn to the low energy limit in which the coherence length tends to zero ($\Delta \to \infty$). In this case the only surviving contribution is $\sigma_n^{(j=1)}$ -- the only one where the cosinus damping factor is absent. 
Note that $T_n^{(j=1)}(b) = T_A^n (b)$.
Removing the second term $(\widetilde{\sigma} \varepsilon)$ in the first factor of (\ref{new1e}) (which corresponds
to no interaction of the projectile since $\delta=1$) we get in this limit
\beq
\label{10e}
\sigma_{pA}^{\psi}(b) = \sigma_{pN}^{\psi} {\sigma_A^{in} (\sigma_{abs}) \over \sigma_{abs}}
\eeq

\noi with $\sigma_{abs} =  \widetilde{\sigma}(1 - \varepsilon ) + \sigma (1 - \delta )$. Since $\delta = 1$ we obtain the conventional 
Glauber formula with a rescattering controlled by $\sigma_{abs} =  \widetilde{\sigma}(1 - \varepsilon )$ -- which is present for $J/\psi$ production ($\varepsilon \sim 0$) and absent for open charm ($\varepsilon = 1$). \par

So far our formulae are valid for mid-rapidities. If we want to consider large positive values of $x_F$, 
we have to take into account that the geometry introduces a correlation between $x_F$ and impact parameter $b$. Indeed, for very peripheral collision (where the projectile goes through the edge of the nucleons) the number of wounded 
nucleons (i.e. cut interactions of the projectile with target nucleus) is small and production at large $x_F$ is favored. In ref. \cite{6r} this correlation was taken into account phenomenologically by introducing a softening factor 
$F_k = (1 -x_1^{\gamma})^{k-1}$. 
Here $x_1$ ($x_2$) is the $x$ variable relative to the projectile (target) with $x_1x_2s = M_{c\overline{c}}^2$ and $x_F = x_1 - x_2$. $k$ is the number of cut interactions
(in Ref. \cite{6r}, $\gamma=2$ for $J/\psi$ production). 
With the factor $F_k$ the spectrum of $J/\psi$ gets softer with increasing $k$. Since in eqs. (\ref{7e}), (\ref{8e}) and (\ref{10e}) we have kept track of the cut interactions it is immediate to introduce the softening factor $F_k$. 
In the low energy limit 
we have to replace $\sigma (\delta -1)$ by $\sigma \delta (1-x_1^{\gamma}) -\sigma= -\sigma x_1^{\gamma}$ $(\delta=1)$.
In this way,
the absorptive cross-section $\sigma_{abs} = 
(1 - \varepsilon) \widetilde{\sigma}$ is replaced by an effective cross-section $\sigma_{eff} = \sigma_{abs} + \sigma x_1^{\gamma}$. Thus, when $x_1 \to 1$ we get rescatterings controlled by a typical hadronic cross-section. It was shown in \cite{6r} that such a large cross-section is needed in order to reproduce the $J/\psi$ suppression at large $x_F$. Note that the same increase of the effective absorptive cross-section as $x_1 \to 1$ occurs for open charm production. 
Likewise, the softening factor has to be introduced in the high energy limit.
However, its effect at the highest available rapidity at RHIC energies ($y=2.2$) is very small and can be neglected.

The extension of the above formalism to $AB$ collisions is straightforward (see for instance Ref. \cite{9r}). 
The formulae are obtained by a convolution
of the ones for $pA$ and $pB$.


\section{Nuclear shadowing}

The above considerations 
--that is to say, the $A^1$ dependence in eqs. (\ref{3e}) and (\ref{9e})--
suggest a phenomenological analysis of $J/\psi$ suppression in $dAu$ collisions at RHIC energies using $\sigma_{abs} = 0$ and taking only into account nuclear shadowing. The latter is particularly important because the data are presented as a ratio $\sigma_{dAu}^{J/\psi }/N_{coll}$ -- rather than a ratio $\sigma_{dAu}^{J/\psi}/ \sigma_{dAu}^{D-Y}$ in which shadowing effects would cancel to a large extent. 

We call shadowing
the mechanisms that makes that
the nuclear structure functions in nuclei are different from the superposition of
those of their constituents nucleons.
Typically,
the shadowing behaves as follows: it increases with decreasing $x$
and decreases with increasing
$Q^2$.
Several explanations to the shadowing have been proposed. We can distinguish between 
two main approaches: models based on multiple scattering \cite{9r}-\cite{FGS} in the rest frame of the nucleus, and models based on 
a frame in which the nucleus is moving fast, and 
where gluon recombination due to the overlap of 
the gluon clouds from different nucleons reduces the gluon density in the nucleus. 

In the rest frame on the nucleus,
nuclear shadowing can be seen as a consequence of multiple scattering
\cite{9r}-\cite{FGS}
the incoming virtual photon splits into a colorless $q\bar{q}$ pair long before
reaching the nucleus, and this dipole interacts with typical hadronic cross sections
which results in absorption.
Multiple scattering can be related to diffraction by means of the
Abramovsky-Gribov-Kancheli (AGK) rules.

On the other hand, another possible approach consists on the
models based on Dokshitzer-Gribov-Lipatov-Altarelli-Parisi (DGLAP) evolution \cite{DGLAP}
like EKS98 \cite{EKS}, nDS \cite{nDS} or HKN \cite{HKN}.
This type of models do not try to address the origin of nuclear shadowing
but study the $Q^2$-evolution of nuclear ratios of parton densities,
\beq
\label{eqeks}
R^A_i (x,Q^2) = \frac{f^A_i (x,Q^2)}{ A f^{nucleon}_i (x,Q^2)}\ , \ \
f_i = q, \bar{q}, g,
\eeq
through the DGLAP evolution equations.
They try to perform for the nuclear case the same program developed for the nucleon:
nuclear ratios are parameterized at some value $Q^2_0 \sim 1 \div 2$
GeV$^2$ which is assumed large enough for perturbative DGLAP evolution to be applied reliably.
These initial parameterizations for every parton density have to cover the full
$x$ range $0 < x < 1$.
Then these initial conditions are evolved through the DGLAP equations towards
larger values of $Q^2$ and compared with experimental data.
From this comparison the initial parameterizations are adjusted.
While DGLAP approaches do not address the fundamental problem of the origin of shadowing, they are of great practical interest. They provide the parton densities required to compute cross sections for observables characterized
by a hard scale for which collinear factorization can be applied,

In the present paper, we are going to compare the data with the results
obtained from \cite{9r} -model based on multiple scattering- and
from EKS98 \cite{EKS} -model based on the GRV LO parton densities and
where the shadowing ratios for each parton type are evolved to LO for
$1.5 < Q < 100$ GeV-.

Following the first approach, 
in order to compute the shadowing we need the total contribution which arises from cutting the two-exchange amplitude in all possible ways (between the amplitudes and the amplitudes themselves in all possible manners). It can be shown that, for purely imaginary amplitudes, this total contribution is identical to minus the contribution from the diffractive cut. Thus 
diffraction becomes linked to the first contribution of the nuclear shadowing. 

In consequence, we apply here a dynamical, non linear shadowing \cite{9r} which is a
 consequence of multiple
scattering, and it
is determined in terms of the diffractive cross-sections. 
It would lead to saturation at $s \rightarrow \infty$.
It is
controlled by triple Pomeron diagrams.
This kind of shadowing gives a positive contribution to the diffractive 
cross-section,
and, on the contrary, its
contribution to the total cross-section is negative. It gives good results
for
the kinematic regions where nuclear DIS data exist \cite{salgadoyo}.
The effect of the shadowing corrections in our approach \cite{9r}-\cite{8r} can be expressed by the factor:
\beq
\label{5ec}
R_{AB}(b) = {\int d^2s \ f_A(s)\ f_B(b-s) \over T_{AB}(b)}
\eeq
where
\beq
\label{6ec}
f_A(b) = {T_A(b) \over 1 + A\ F(s) \ T_A(b)} \ .
\eeq
Here the function $F(s)$ is given by the integral of the ratio
of the triple Pomeron
cross-section $d^2\sigma^{PPP}/dy dt$ at $t = 0$ to the single Pomeron exchange
cross-section $\sigma_p(s)$~:
\beq
\label{7ec}
\left . F(s,y) = 4 \pi \int_{y_{min}}^{y_{max}} dy \ {1 \over \sigma_p(s)} \ {d^2
\sigma^{PPP} \over dy\ dt} \right |_{t=0} = C \left [ \exp \left (
\Delta y_{max}\right )
- \exp \left ( \Delta y_{min}\right ) \right ] \eeq
with $y = \ell n (s/M^2)$, where $M^2$ is the squared mass of
the diffractive
system. For a particle produced at $y_{cm} = 0$, $y_{max} = {1 \over 2} \ell n
(s/m_T)^2$ and $y_{min} = \ell n (R_A m_N/\sqrt{3})$. $\Delta =
\alpha_P(0) - 1 = 0.13$
and $C$ is a constant proportional to the triple Pomeron coupling. $R_A$
is the nuclear radius, $T_A(b)$ the nuclear profile function and
$T_{AB}(b) = \int
d^2s T_A(s) T_B(b-s)$.
The denominator in Eq. (\ref{5ec}) correspond to the sum of
all ``fan''
diagrams with Pomeron branchings (generalized Schwimmer model \cite{Schwi}).
Note that shadowing corrections to inclusive spectra are not specific to soft
processes. The triple Pomeron terms described above are also
responsible for shadowing
in hard processes.

We have then for the suppression of hadron $h$ production in $AB$ collisions
\beq
\label{11e}
S^h(b,s,y) = {1 \over 1 + A F_h (y_A) T_A (s)}\ {1 \over 1 + BF_h(y_B) T_B (b-s)}
\eeq

\noi where $F_h (y_A) = C [\exp (\Delta y_{max}) - \exp (\Delta y_{min})]$, $y_{min} = \ell n (R_A m_N /\sqrt{3})$, $\Delta = 0.13$ and $C = 0.31$~ fm$^2$. $y_{max}$ depends on the rapidity of $h$~: $ y_{max}^A = {1
 \over 2} \ell n (s/m_T^2) \pm y$ with the $+$ ($-$) sign if $h$ is produced in the hemisphere of nucleus $B(A)$. $m_T$ is the transverse mass of $h$. 
For charged particles we use $m_T = 0.4$~GeV and for $J/\psi$ $m_T = 3.1$~GeV. 
For $dAu$ collisions we just take the first factor of (\ref{11e}), which corresponds to $pA$ collisions. 
With $\sigma_{abs} = 0$ this suppression equals the total $J/\psi$ suppression, 
since no {\it hot} -final state interactions- effects are expected in this case. 

We are going to compare the results of our model with the ones obtained from the EKS98
parameterization. In order to do that, we have applied the same procedure as the
one introduced by R. Vogt in Ref. \cite{Vogt}.
The spatial dependence can be parameterized 
assuming that shadowing
is proportional to the local density, $\rho_A(s)$,
\beq
R^A_i (x,Q^2,\vec{r},z)=1+N_{WS}[R^A_i (x,Q^2)-1]\frac{\rho_A(s)}{\rho_0}\ ,
\eeq
where
$\vec r$ and $z$ are the transverse and longitudinal location in position
space, with $s=\sqrt{|\vec r|^2+z^2}$,
$\rho_A(s)$ corresponds
to the Woods-Saxon distribution for the nucleon density in the nucleus,
$\rho_0$ is the central density, given by the normalization
$\int d^2r dz \rho_A(s) = A$
and
$R^A_i (x,Q^2)$ is the shadowing function from EKS98 as defined by eq. (\ref{eqeks}).
$N_{WS}$ is chosen so that
$(1/A) \int d^2r \int dz \rho_A(s) R^A_i (x,Q^2,\vec{r},z) = R^A_i (x,Q^2)$
In the small $x$ limit, the above formulae transforms into
\beq
R^A_i (x,Q^2,\vec{r},z)=1+N_{\rho}[R^A_i (x,Q^2)-1]
\frac{\int dz \rho_A(\vec r,z)}{\int dz \rho_A(0,z)}\ .
\eeq
The integral over z includes the material traversed by the
incident nucleon, 
so we are considering that the
incident parton interacts coherently with all the target partons along its pathlength.\\
The normalization requires again
$(1/A) \int d^2r \int dz \rho_A(s) R^A_i (x,Q^2,\vec{r},z) = R^A_i (x,Q^2)$.
We do not have included the nuclear absorption, since we consider $\sigma_{abs}=0$.

\section{Results}

Our results 
of the nuclear modification factor for the $J/\psi$
--that is, the $AB/pp$ ratio for the $J/\psi$,
$\frac{dN_{AB}^{J/\psi}/dy}{dN_{pp}^{J/\psi}/dy}$, at a given centrality,
normalized by the number of collisions $N_{coll}^{AB}$ at this centrality, $R_{A
B}=\frac{dN_{AB}^{J/\psi}/dy}{N_{coll}^{AB}
dN_{pp}^{J/\psi}/dy}$--
 for $dAu$ collisions, compared with
PHENIX data \cite{2r}, are presented in Fig. 1. 
They are also compared with the results obtained in the framework of the EKS98 model \cite{EKS,Vogt}.
 
The agreement between theory and experiment is quite good in both cases. 
Note that the increase of $J/\psi$ suppression with rapidity is entirely due to the corresponding increase of 
shadowing.

The situation is different in $Au$ $Au$ collisions. Indeed, it can be checked with our formulae that shadowing effects are, in this case, practically independent of rapidity. This is due to the opposite signs of $y$ in $y_{max}^A$ and $y_{max}^B$. 
As a consequence, the $J/\psi$ suppression in $Au$ $Au$ in a comovers approach 
-- a charmonium state produced in a primary nucleon-nucleon collision is dissociated through interactions with 
the dense
medium subsequently formed in the collision --
is maximal at $y = 0$, 
where the comovers density is maximal,
and decreases when we move to positive or negative rapidities. \par
Moreover, this is the case in any model that 
considers the suppression of the $J/\psi$ dependent on the medium density, that is,
such dissociation could occur in a confined \cite{PRL} as well as in a deconfined medium \cite{Satz}.

One possibility to reverse this tendency should be the recombination of $c$-$\overline{c}$ pairs into $J/\psi$
\cite{recom}. In this case,
in the hadronization of the quark-gluon plasma, 
charmonium formation can occur by binding of a $c$ with a $\overline{c}$ from different 
nucleon-nucleon collisions, as well as from the same. If the total number of available $c$-$\overline{c}$ 
pairs considerably exceeds their thermal abundance, statistical recombination enhances 
hidden relative to open charm production, as compared to hadron-hadron collisions.
This possibility has been neglected in our model. It would correspond to the introduction of a gain term in 
differential equations which govern the final state interactions. We will develop this possibility in our future 
work.

The results for $Au$ $Au$ collisions at RHIC at $y = 0$ with $\sigma_{abs} = 0$ and $\sigma_{co} = 0.65$~mb (the same comovers interaction cross-section obtained from the CERN data) were given in \cite{9r}. We present them again in Fig.~2 and compare them with the 
PHENIX data \cite{10r}. The corresponding results at $y = 1.7$ are also presented. 
The agreement with the data is considerably improved by taking $\sigma_{abs} = 0$. 
Nevertheless, our results are still somewhat lower than the 
data at $y =0$, while there are higher at $y=1.7$. 
This shows that in the present version of our model the rapidity dependence of the suppression is not reproduced.
Moreover, it
is in contradiction with the present
data, as it has been indicated above: the $J/\psi$ suppression in $Au$ $Au$ is maximal at $y = 0$,
where the medium density is maximal,
and decreases when we move to positive or negative rapidities.
\par

Note that 
$\sigma_{co}$ is the comover cross-section properly averaged over the momenta of the colliding particles and over 
the collision time.
A large contribution to this interaction comes from the few first fm/c
after the collision
-- where the system is in a pre-hadronic or partonic stage. 
Actually, Brodsky and Mueller \cite{17r} introduced the comover
interaction as a
coalescence phenomenon at the partonic level. In view of that, there is no
precise connection
between $\sigma_{co}$ and the physical $J/\psi - \pi$ or $J/\psi - N$
cross-section, and
$\sigma_{co}$ has to be considered as a free parameter. 
Since it is a cross-section near threshold, we do not expect a substantial energy variation. 
Because of this, we have used the same comovers interaction cross-section obtained from the CERN data.
However, this cross-section does not have to be identical at CERN and 
RHIC -- since the momentum distribution of the comovers can change with energy. 
Unfortunately we are unable to evaluate such an effect. 
Another source of uncertainty in our calculation resides in the comovers density. 
%
As an illustration, we show in Fig.~2 the effect on the $J/\psi$ suppression at $y=0$ resulting from a 20$\%$ decrease in this density.

More important, theoretical calculations \cite{Kai} show that the energy dependence of the $J/\psi$-hadron cross-section increases very fast with increasing energy near threshold. 
Because of that, the decrease of the comovers density at forward rapidities might be (over)compensated by an increase 
of the comovers cross-section $\sigma_{co}$. In this case, the maximum of the $J/\psi$ suppression may take place
at a value of the rapidity different from $y=0$ where the comovers density is maximal. This point, 
as well as the possibility of recombination,  
are at present under 
investigation\footnote{We thank A. B. Kaidalov for interesting suggestions on this point.}.

\section{Conclusions}

In conclusion, we have presented theoretical arguments supporting the idea that, at high energy,
 nuclear absorption of the $J/\psi$ vanishes ($\sigma_{abs} = 0$). Only standard nuclear effects (shadowing) are present. This produces an increase of the $J/\psi$ suppression with increasing rapidity in $dAu$ collisions. The results at $\sqrt{s} = 200$~GeV energies can be described in this way. In $Au$ $Au$ collisions the shadowing effect is practically independent of rapidity and the $J/\psi$ suppression in 
the comovers interaction model presented here
decreases when moving away from mid-rapidity, due to the corresponding decrease of the comovers density.\par

It is a pleasure to thank Olivier Drapier, Frederic Fleuret, Raphael Granier de Cassagnac, Ermias T. Atomssa and Andry 
Rakotozafindrabe for useful comments and inestimable help during the writing of this manuscript. 
E. G. F. wants also to thank the program Ram\'on y Cajal of the Ministerio de Educaci\'on y Ciencia of Spain and 
the Laboratoire Leprince-Ringuet of Ecole Polytechnique, CNRS-IN2P3, Palaiseau, France, where part of this work was developed.  

We also thank N. Armesto, M. A. Braun, C. Pajares and C. A. Salgado 
for enlighting discussions on the theoretical aspects of this work.

\newpage
\begin{figure}[hbp]
\begin{center}
\vskip 2cm
\epsfxsize=6in
\epsfysize=2in
\epsffile[72 408 558 558]{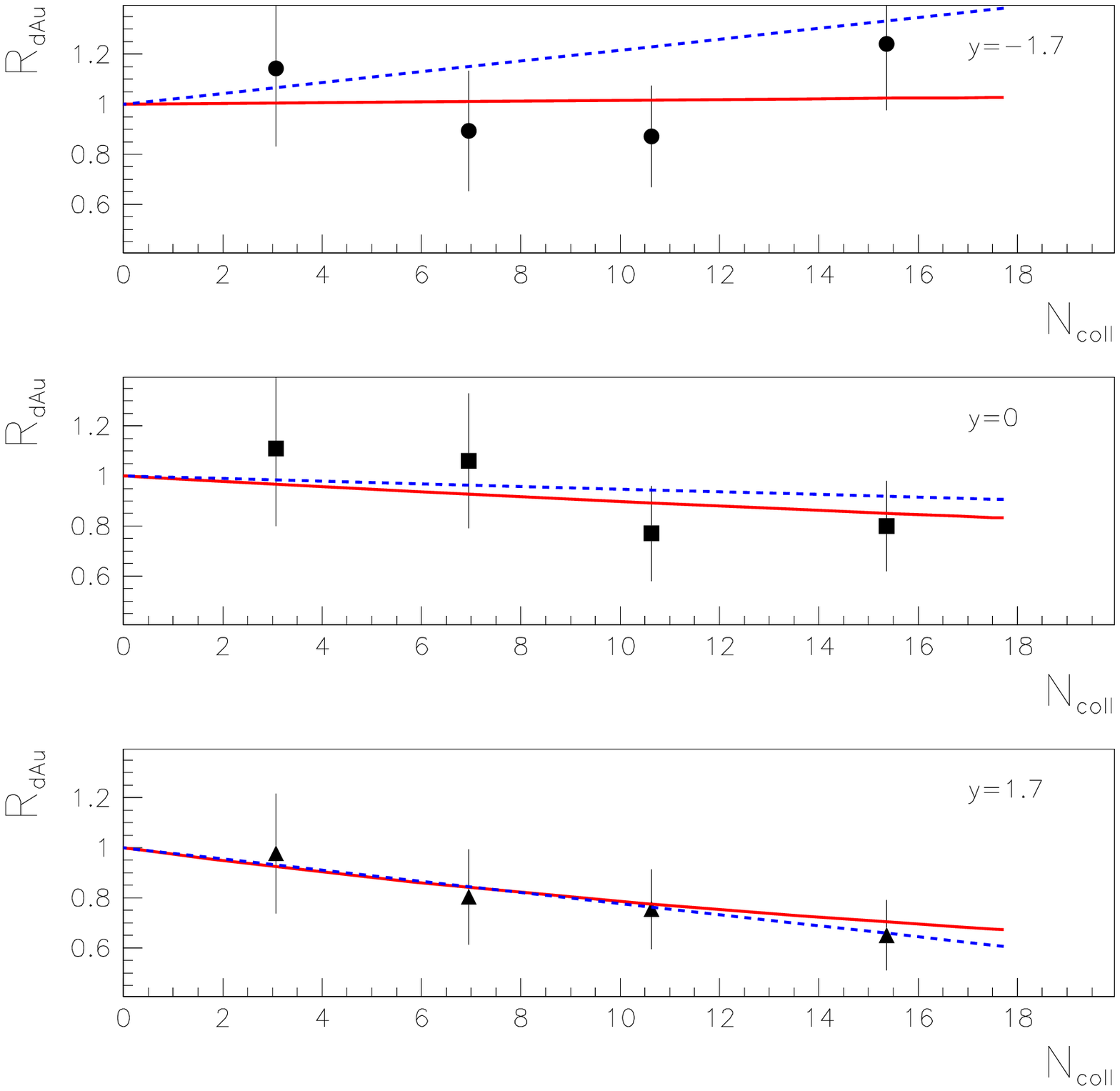}
\end{center}
\vskip 8cm
\caption{(Color online) Results on $J/\psi$ suppression for $dAu$ collisions
vs. the number of collisions
in 3 different rapidity regions. We take into account nuclear shadowing alone ($\sigma_{abs}=0$).
Data are from PHENIX. Continuous line: our results from Pomeron shadowing, discontinuous line: results 
from EKS98 shadowing.}
\label{fig1}
\end{figure}

\newpage
\begin{figure}[ht]
\begin{center}
\psfig{figure=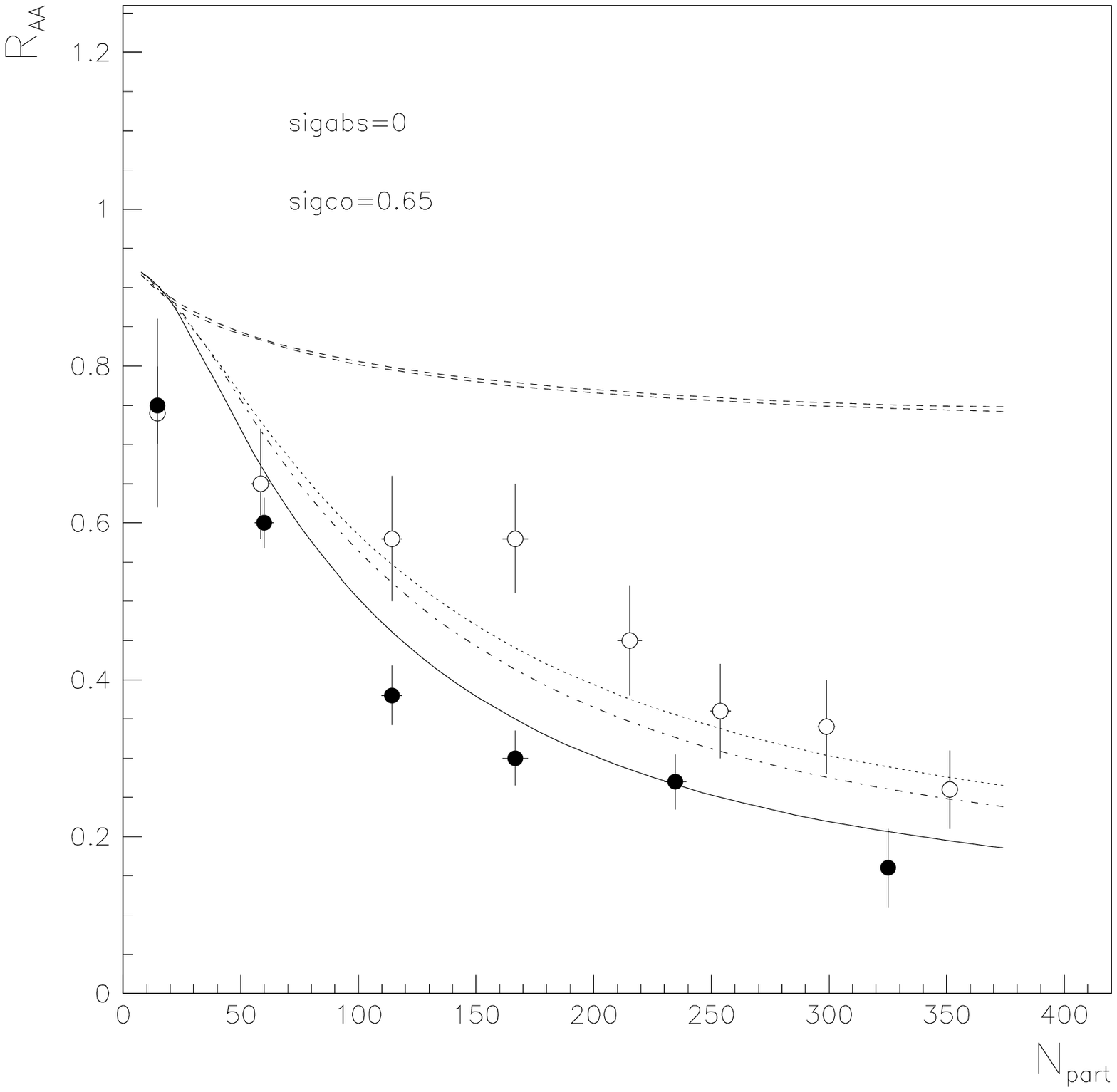,height=7.0in}
\end{center}
\vskip -4cm
\caption{Results on $J/\psi$ suppression for $AuAu$ collisions
vs. the number of participants at $y=0$ (solid line) and $y=1.7$ (dotted line). 
The dashed-dotted line is the result at $y=0$ obtained by decreasing the comovers density 
by 20$\%$ (see main text).
The dashed lines correspond to the suppression
from shadowing alone.
Data are from PHENIX (black points: data at $y=1.7$, open points: data at $y=0$).}
\label{fig2}
\end{figure}


\newpage
\begin{thebibliography}{99}
\bibitem{1r} NA50 Collaboration: L. Ramello,
Nucl. Phys. {\bf A715}, 243c (2003). \\
B. Alessandro et al., NA50 Collaboration, Eur. Phys. J. {\bf C39}, 
335 (2005);
CERN-PH-EP-2006-018, to appear
in Eur. Phys. J.
 

\bibitem{2r} PHENIX Collaboration, Phys. Rev. Lett. {\bf 96}, 012304 (2006).

\bibitem{3r} R. J. Glauber in Lectures in Theoretical Physics, ed. ;\\
W. E. Britten (Interscience, New York 1959) ;\\
A. G. Sitenko, Ukr. Phys. J. {\bf 4}, 152 (1959) ;\\
V. N. Gribov, Sov. Phys. JETP {\bf 29}, 483 (1969)~; ibid {\bf 30}, 709 (1970).

\bibitem{7r} M. A. Braun, C. Pajares, C. A. Salgado, N. Armesto and A. Capella, Nucl. Phys. {\bf B509}, 357 (1998);\\
C. A. Salgado, Phys. Lett. {\bf B521}, 211 (2001);\\
A. Capella, nucl-th/0207049. 
 
\bibitem{4r} R. Blankenbecler, A. Capella, C. Pajares, A. V. Ramallo and J. Tran Thanh Van, Phys. Lett. {\bf B107}, 106 (1981).

\bibitem{5r} B. Z. Kopeliovich, A. V. Tarasov and J. H\"ufner, Nucl. Phys. {\bf A696}, 669 (2001).

\bibitem{6r} K. Boreskov, A. Capella, A. Kaidalov and J. Tran Thanh Van, Phys. Rev. {\bf D47}, 919 (1993).

\bibitem{9r} A. Capella and E. G. Ferreiro, Eur. Phys. J. {\bf C42}, 419 (2005).

\bibitem{8r} A. Capella, A. Kaidalov and J. Tran Thanh Van,
Heavy Ion
Phys. {\bf 9}, 169 (1999).

\bibitem{FGS} L. Frankfurt, V. Guzey and M. Strikman, Phys. Rev. {\bf D71}, 054001 (2005).

\bibitem{DGLAP} Y. L. Dokshitzer, Sov. Phys. JETP {\bf 46}, 641 (1977); 
V. N. Gribov and L. N. Lipatov, Sov. J. Nucl. Phys. {\bf 15}, 438 (1972);
G. Altarelli and G. Parisi, Nucl. Phys. {\bf B126}, 298 (1977).

\bibitem{EKS} K. J. Eskola, V. J. Kolhinen and C. A. Salgado, Eur. Phys. J. {\bf C9}, 61 (1999).

\bibitem{nDS} D. de Florian and R. Sassot, Phys. Rev. {\bf D69}, 074028 (2004). 

\bibitem{HKN} 
M. Hirai, S. Kumano and T.-H. Nagai, 
Phys. Rev. {\bf C70}, 044905 (2004);
e-Print: arXiv:0709.3038.

\bibitem{salgadoyo} A. Capella, E. G. Ferreiro, C. A. Salgado and A. B. Kaidalov, Phys. Rev. {\bf D63},
054010 (2001); Nucl. Phys. {\bf B593}, 336 (2001).

\bibitem{Schwi} A. Schwimmer, Nucl. Phys. {\bf B94}, 445 (1975).

\bibitem{Vogt} R. Vogt, Acta Phys. Hung. {\bf A25}, 97 (2006); Phys. Rev. {\bf C71}, 054902 (2005).

\bibitem{PRL} A. Capella, E. G. Ferreiro and A. B. Kaidalov, Phys. Rev. Lett. {\bf 85}, 2080 (2000).

\bibitem{Satz} D. Kharzeev and H. Satz, Phys. Lett. {\bf B334}, 155 (1994). 

\bibitem{recom} P. Braun-Munzinger and J. Stachel, Nucl. Phys. {\bf A690}, 119 (2001); R. L. Thews et al., 
Phys. Rev. {\bf C63}, 054905 (2001);
L. Grandchamp and R. Rapp, Nucl. Phys. {\bf A709}, 415 (2002).

\bibitem{10r} PHENIX Collaboration, Phys. Rev. {\bf C69}, 014901 (2004); nucl-ex/0611020\\
F. Fleuret for PHENIX Collaboration, proceedings of Conference on Quark and Nuclear Physics, Madrid, Spain, June 2006;\\
R. Granier de Cassagnac, proceedings of International Conference on Hard Probes of High Energy Nuclear Collisions, 
Pacific Grove, California, June 2006.

\bibitem{17r} S. J. Brodsky and A. H. Mueller, Phys. Lett. {\bf B206}, 685
(1988).

\bibitem{Kai} G. Bhanot and M. E. Peskin, Nucl. Phys. {\bf B156}, 391 (1979).

\end{thebibliography}
\end{document}